\documentclass[10pt,prd,aps,twocolumn,nofootinbib,superscriptaddress,eqsecnum,floatfix,preprintnumbers,amsmath,amssymb,longbibliography]{revtex4-2}

\usepackage{xcolor}
\usepackage{amsmath}
\usepackage{graphicx}
\usepackage{epsfig}
\usepackage[dvipsnames]{xcolor}
\usepackage[dvipsnames]{xcolor}
\usepackage[utf8]{inputenc}
\usepackage{stmaryrd}
\usepackage{mathrsfs}
\usepackage{mathalfa}
\usepackage{accents}
\usepackage{enumitem}
\usepackage[normalem]{ulem}

\usepackage[colorlinks=true,
            citecolor=green,
            linkcolor=red,
            filecolor=cyan,
            urlcolor=magenta,
            backref=false]{hyperref}

\newcommand{\pa}{\partial}

\newcommand{\be}{\begin{equation}}
\newcommand{\ee}{\end{equation}}
\newcommand{\bea}{\begin{eqnarray}}
\newcommand{\eea}{\end{eqnarray}}
\newcommand{\ba}{\begin{equation}\begin{aligned}}
\newcommand{\ea}{\end{aligned}\end{equation}}
\newcommand{\beg}{\begin{gather*}}
\newcommand{\eng}{\end{gather*}}
\newcommand{\hh}{,\hspace{0.5cm}}
\newcommand{\hhh}{,\hspace{0.2cm}}

\newcommand{\n}[1]{\label{#1}}

\newcommand{\CAL}{\mathcal}

\begin{document}

\title{Charged Black Holes in Quasi-Topological Gravity Coupled to Born-Infeld Nonlinear Electrodynamics}

\author{Jose Pinedo Soto}
\email[]{pinedoso@ualberta.ca}
\affiliation{Theoretical Physics Institute, Department of Physics,
University of Alberta,\\
Edmonton, Alberta, T6G 2E1, Canada
}

\author{Valeri P. Frolov}
\email[]{vfrolov@ualberta.ca}
\affiliation{Theoretical Physics Institute, Department of Physics,
University of Alberta,\\
Edmonton, Alberta, T6G 2E1, Canada
}
\affiliation{Kobayashi-Maskawa Institute for the Origin of Particles and the Universe (KMI),\\
Nagoya University, Nagoya 464-8602, Japan}

\begin{abstract}
We construct static, spherically symmetric black hole solutions in quasi-topological gravity (QTG) coupled to Born-Infeld nonlinear electrodynamics. Starting from the spherically reduced action, we derive closed-form expressions for the electric field, the nonlinear Lagrangian, and the metric function, the latter involving hypergeometric functions. We consider specific versions of QTG in which vacuum black holes are regular, and show that, for some of these models, charged black holes develop a curvature singularity at a finite radius in their interior. In contrast, in models such as a Born-Infeld-type QTG, charged black holes remain regular. In this case, however, the de Sitter core of the neutral solution is replaced by an anti-de Sitter core. We also discuss several limiting regimes of these solutions.
\end{abstract}
\maketitle

\section{Introduction}

It is now well understood that General Relativity (GR), despite its remarkable successes, remains incomplete: in particular, it predicts curvature singularities at the centres of black holes \cite{Misner:1973prb}, signalling a breakdown of the theory. Motivated by this limitation, a long-standing effort has been devoted to constructing regular (i.e., nonsingular) black hole models.

The first such model was the Bardeen black hole \cite{Bardeen1968}, later reinterpreted as the gravitational field of a nonlinear magnetic monopole \cite{AYONBEATO_Bardeen}.
One of the earliest works in which regular black holes, their quantum evaporation, and their conformal diagrams were analysed together is \cite{FrVi}.
A characteristic feature of regular black holes is the presence of a de~Sitter-like core in their interior, replacing the near-singularity region of the Schwarzschild solution.
The idea of such a de~Sitter core was first proposed by Gliner \cite{Gliner:1966cgu} and later developed extensively by Dymnikova \cite{Dymnikova:1992ux,Dymnikova_2002,Dymnikova_2004}.
An appealing property of regular black hole models is that they naturally allow the discussion of scenarios in which new universes may form inside black holes \cite{FMM_1,FMM_2,FrBa}.

A wide variety of regular black hole solutions have since been proposed, including the Hayward metric \cite{Hayward}, later generalized in \cite{FrolovMetric}. Other examples include the Simpson-Visser construction \cite{SimpsonVisser}, models in quadratic gravity \cite{Berej2006}, $2D$ dilaton gravity \cite{PhysRevD:lemos,PhysRevResearch:frolov,PhysRevD:Kunstatter}, and non-polynomial gravity \cite{bueno2025nonpolynomial,collieaux2018}. This list is far from exhaustive; for comprehensive reviews, see \cite{nonsingularparadigmblackhole,Bambi2023,Lan2023RBHreview}.

A particularly interesting recent development in the theory of regular black holes arises in the context of quasi-topological gravity (QTG)
\cite{Bueno:2020gq,Bueno:2019ltp,Bueno:2022res,Oliva_2010,Hennigar:2017ego,Myers:2010ru,Moreno:2023rfl,QT_BH,rbh_pfkz,PinedoSoto:2025hel, Bueno:2025tli, Bueno:2025zaj, Frolov:2025ddw, Bueno:2026dln,Sueto:2026epz,ALKAC2025170133,borissova20262dgeneraliseddilatontheories,borissova2026gttgrr1blackhole,borissova2026regularblackholespure},.
In this approach, formulated in spacetime dimensions $D\ge 5$, an infinite series of higher-curvature terms is added to the Einstein-Hilbert action. Remarkably, for special combinations of these invariants, the spherically symmetric field equations do not contain derivatives higher than second order. This structure provides significant freedom in shaping the gravitational action and enables the construction of regular black hole solutions.

However, these regular metrics typically possess an inner Cauchy horizon, which often introduces dynamical instabilities \cite{bonanno2025}. A similar problem arises in GR for charged Reissner-Nordström black holes, where the Cauchy horizon becomes singular when perturbed—a phenomenon known as mass inflation \cite{PoissonIsrael_CH,PoissonIsrael_CH2,Ori_CH}.
The problem of mass inflation within the framework of QTG was recently analyzed in \cite{Frolov:2026rcm}, where it was shown that, for regular macroscopic black holes, the region in which one may expect a blow-up of curvature is shifted to trans-Planckian scales.

Interestingly, nonlinear electrodynamics can sometimes mitigate or remove these instabilities \cite{hale2025CauchyBI,hale2025rotextBI}.
Nonlinear electrodynamics has also long been explored as a mechanism for eliminating singularities altogether \cite{SorokinNED,SingularitiesNED,PRL:Garcia,PRD:Bronnikov,Dymnikova_2004,Kubiznak:2022vft}.
These developments motivate the investigation of regular black holes in modified theories of gravity coupled to nonlinear electrodynamics. It is worth noting, however, that such constructions often require a fine-tuning of the black hole charge and mass parameters.

In this paper, we study charged black holes within the framework of QTG. Specifically, we consider particular versions of QTG in which neutral vacuum black holes are regular, and investigate how these solutions are modified in the presence of charge. To this end, we couple QTG to nonlinear electrodynamics. In Sec.~\ref{sec2}, we derive the spherically reduced action for this system. It is characterized by two arbitrary functions: $h(p)$, where $p$ is the primary curvature invariant associated with the QTG sector, and the Lagrangian density $\mathcal{L}(\mathcal{E})$ for the electric field $\mathcal{E}$ in the nonlinear electrodynamics sector.

The model is characterized by two dimensional parameters, $\ell$ and $b$. The parameter $\ell$, entering the QTG action, sets the length scale at which higher-curvature corrections become significant and modify Einstein gravity. The second parameter, $b$, determines the electric field scale at which nonlinear electrodynamic effects become important.
One can introduce the dimensionless combination $\varkappa \ell^{2} b^{2}$, where $\varkappa$ is the gravitational coupling constant. This parameter controls the relative importance of the two cutoff scales. In what follows, we treat this parameter as arbitrary and use it to explore different limiting regimes.

In order to derive the field equations for the system and construct their solutions, we employ the following general form of the spherically symmetric static metric
\be \n{SSM}
ds^2=-N^2 f dt^2+\dfrac{dr^2}{f}+r^2 d\Omega_{D-2}^2\, ,
\ee
where $N$ and $f$ are arbitrary functions of $r$, and $d\Omega_{D-2}^2$ denotes the metric on the $(D-2)$-dimensional unit sphere. The resulting field equations imply that $N$ is constant, and it can be set to unity without loss of generality. Further analysis shows that the general solution of the field equations can be written in an integral form. In addition to integration, this construction requires only algebraic operations. Specifically, one must determine the inverse function $p=p(h)$ of $h=h(p)$, and express the electric field $\mathcal{E}$ contribution in terms of the quanitity $\mathcal{D}$ which is defined by $\mathcal{D}=d\mathcal{L}/d\mathcal{E}$. 

In Sec.~\ref{sec3}, we consider a well-known special case of nonlinear electrodynamics, namely the Born–Infeld theory. In this case, the inversion of the relation $\mathcal{D}=\mathcal{D}(\mathcal{E})$ can be performed analytically, and the required integrals can be evaluated explicitly in terms of hypergeometric functions.
In this section we also study general properties of the obtained solutions.

In Sec.~\ref{sec4}, we consider two specific realizations of QTG models: the Hayward-type and the Born–Infeld-type. In both cases, neutral vacuum black holes are known to be regular. We show, however, that the corresponding charged solutions exhibit a qualitatively different behaviour.
In the presence of charge, Hayward-type black holes can develop a curvature singularity at a finite radius within their interior, whereas Born–Infeld-type solutions remain regular for all values of the parameters. Notably, in the latter case the internal structure differs from that of the vacuum solutions: instead of a de Sitter–like core, the geometry approaches an anti–de Sitter–like core inside the black hole.

Section~\ref{sec_extremal} is devoted to extremally charged black holes in QTG, including the construction of the charge-to-mass ratio for extremal configurations. Special limiting regimes of the general solutions are discussed in Sec.~\ref{sec_specialcases}, while Sec.~\ref{sec7} contains a summary and discussion of the results.

Throughout this paper, we adopt the standard MTW sign and unit conventions \cite{Misner:1973prb} for gravitational and electromagnetic equations, generalized to higher dimensions.

\section{quasi-topological gravity coupled to nonlinear electrodynamics}\label{sec2}

\subsection{Action}

We write the action of quasi-topological gravity coupled to nonlinear electrodynamics in the form
\be \n{SS}
S[\,g,\,A]=S_{QT}+S_A\, .
\ee
The action of quasi-topological gravity is
\be \n{QTA}
S_{QT}=\frac{1}{2\varkappa}\,\int d^D x \,\sqrt{-g}\,[ R + \sum_j \alpha_j \ell^{2(j-1)} Z_j ]\, .
\ee
Here $\varkappa=8\pi G^{D}$ and $G^D$ is the $D-$dimensional gravitational coupling constant.

The quantities $Z_j$, which enter $S_{QT}$, are specially constructed scalar invariants, which are polynomials of order $j$ in curvature. The explicit form of the $Z_j$ and recursive relations for their construction can be found in \cite{Bueno:2020gq}. The parameter $\ell$, which has the dimension of length, plays the role of the fundamental length and specifies the scale where the higher in curvature terms are important. The dimensionless parameters $\alpha_j$ specify the model.

We will write the action of our nonlinear electrodynamics model in the form
\begin{equation}
\begin{split}
&S_A[A_{\mu}]= \int d^D x \, \sqrt{-g} {L}(\mathcal{F})\, \\
&\mathcal{F}=-\frac{1}{2}F_{\mu\nu}F^{\mu\nu} \, ,\quad
F_{\mu\nu}=A_{\mu;\nu}-A_{\nu;\mu}\, .
\end{split}
\end{equation}
Here ${L}$ is a function of the invariant $\mathcal{F}$
\footnote{
We do not include the other quadratic in the field strength invariant since it vanishes for the static spherically symmetric field, which we shall consider in the paper.} and a semicolon denotes a covariant derivative. We assume that for small $\mathcal{F}$ the function ${L}(\mathcal{F})$ has the expansion
\begin{equation}\label{PPPP}
L(\mathcal{F})=\frac{1}{2}\mathcal{F}+...\, .
\end{equation}
The condition \eqref{PPPP} guarantees that in the weak field limit, the nonlinear electrodynamics model reduces to the standard Maxwell theory.

\subsection{Spherical reduction}

We focus on static, spherically symmetric solutions of the field equations derived from the action \eqref{SS}. The corresponding metric \eqref{SSM} is given by
\be\n{MMM}
ds^2=-N^2 f dt^2+\dfrac{dr^2}{f}+r^2 d\Omega_{D-2}^2\, ,
\ee
where $N$ and $f$ are arbitrary functions of $r$.
Similarly, the vector potential for a static spherically symmetric electric field is
\be \n{Apot}
A_{\mu}=\varphi(r) \delta_{\mu}^t\, ,
\ee
We denote
\be
\CAL{E}=\dfrac{d\varphi}{dr} \, .
\ee
Then one has
\begin{equation}
\mathcal{F}=\frac{\CAL{E}^2}{N^2}\, .
\end{equation}
Here and later, we use a prime for the radial derivative: $(\ldots)'=d/dr(\ldots)$.

There exist two ways to obtain the spherically reduced equations for the metric and the electromagnetic field:
\begin{itemize}
\item Perform variation of the action \eqref{SS} with respect to $g_{\mu\nu}$ and $A_{\mu}$ and after this, use the relations \eqref{MMM} and \eqref{Apot}.

\item Insert the ansatz \eqref{MMM}-\eqref{Apot} directly into the action \eqref{SS} and perform variations over the reduced variables $f$, $N$ and $\varphi$.
\end{itemize}

The equivalence of these
two approaches for the covariant action follows from general results proved in \cite{Fels:2001rv,Anderson:1999cn}. We shall use the second option in which the reduced action for the quasi-topological gravity is known explicitly \cite{QT_BH}.
The spherically reduced form of the action \eqref{SS} is
\begin{equation}\n{RED}
\begin{split}
S[N,f,\varphi]
&=\dfrac{\Omega_{D-2}}{2\varkappa}\int dt dr \bigg[(D-2) N'r^{D-1}h(p) \\
&+2 N \varkappa r^{D-2}L(\CAL{F})\bigg]\, ,
\end{split}
\end{equation}
where $\Omega_{D-2}$ represents the surface area of the unit $(D-2)$-sphere
\begin{equation}
\Omega_{D-2}
= \frac{2 \pi^{\frac{D-1}{2}}}
{\Gamma\!\left(\frac{D-1}{2}\right)}.
\end{equation}

In particular:
\be
\Omega_2 = 4\pi\hhh
\Omega_3 = 2\pi^2\hhh
\Omega_4 = \frac{8\pi^2}{3}.
\ee

The first term in the square brackets is the contribution of quasi-topological gravity. It was derived in \cite{QT_BH} for $D\ge 5$, but it has a well-defined sense in $D=4$. The parameter $p$ is a basic curvature invariant of the metric \eqref{MMM}\footnote{For more details, see e.g. \cite{rbh_pfkz}.}
\be
p=\dfrac{1-f}{r^2}\, .
\ee
The function $h(p)$ depends on the choice of the constants  $\alpha_j$ in \eqref{QTA}.

The second term in the square brackets gives the reduced nonlinear electrodynamic action. Let us emphasize that this part of the action does not depend on the metric function $f(r)$. This happens for two reasons:
\begin{itemize}
\item The determinant of the metric in the integral in \eqref{QTA} does not depend on $f(r)$.

\item The electric field invariant $\CAL{F}$ does not have any dependence on $f(r)$ either.
\end{itemize}

\subsection{Field equations}

The variation of the  reduced action \eqref{RED} with respect to $f$, $\varphi$ and $N$ gives the following set of reduced field equations
\begin{equation}\label{SEQ}
    \begin{split}
&N'=0\, ,\\
&\big(r^{D-2}\dfrac{1}{N} \CAL{E}\dfrac{\pa{L}}{\pa{\CAL{F}}}\big)'=0\, ,\\
&(r^{D-1}h(p))' + \dfrac{2\varkappa}{D-2} r^{D-2} \Big({L} - 2\CAL{F}\frac{\pa{L}}{\pa\mathcal{F}} \Big) = 0\, .
    \end{split}
\end{equation}

The first equation shows that $N=\mathrm{const}$. By simple rescaling of time coordinate $t$ we put $N=1$. Note that for $N=1$ one has $\CAL{F}=\CAL{E}^2$. We denote
\be
\CAL{L}(\CAL{E})=L(\CAL{F})|_{N=1}\, .
\ee
Then one has
\be
\dfrac{\pa{L}}{\pa{\CAL{F}}}\Big|_{N=1}=\dfrac{1}{2\CAL{E}}
\CAL{D}\hh \CAL{D}=
\dfrac{\pa{\CAL{L}}}{\pa{\CAL{E}}}\, ,
\ee
and the second equation in \eqref{SEQ} takes the form
\be
\Big(r^{D-2} \CAL{D} \Big)'=0\, .
\ee
This equation contains only the electric field $\CAL{E}$ and does not contain the metric function. In the presence of the point-like charge
it gives
\begin{equation}\label{eq_ED3}
r^{D-2} \CAL{D}=Q\, .
\end{equation}
Here $Q$ is an integration constant
 which has the meaning of electric charge. We choose its normalization so that the flux of the field $\CAL{D}$ over a sphere $S^{D-2}$ surrounding the charge is
\be
\dfrac{1}{\Omega_{D-2}}\int_{S^{D-2}} d^{D-2}\sigma\, \CAL{D} =  Q\, ,
\ee
where $\Omega_{D-2}$ is the surface area of a unit $(D-2)-$dimensional sphere.
In the weak field approximation $\CAL{D}=\CAL{E}$ and one gets
\be
\CAL{E}=\dfrac{Q}{r^{D-2}}\, .
\ee
We use the relation \eqref{eq_ED3} to simplify the last term in the third equation in \eqref{SEQ}. For $N=1$ this equation takes the form
\begin{equation} \label{rrff}
\begin{split}
&(r^{D-1}h(p))'=-U\, ,\\
&U=\dfrac{2\varkappa}{D-2} \Big[r^{D-2} {L}- Q \CAL{E}
\Big] \, .
\end{split}
\end{equation}

Let us note that equation \eqref{eq_ED3}  can be used to find $\CAL{E}$ as a function of $r$, $\CAL{E}=\CAL{E}(r)$. Substituting this expression in $U$, one can find it as a function of $r$, $U=U(r)$. Equation \eqref{rrff} implies
\be \n{hhpp}
h(p)=\dfrac{1}{r^{D-1}}(Z+\mu)\hh
Z=\int^\infty_{r} U(r') dr'\, .
\ee
Here $\mu$ is an integration constant
related to the mass $M$ measured at infinity as follows
\be \label{eqmu}
\mu=\dfrac{2\varkappa M}{(D-2)\Omega_{D-2}}\, .
\ee

One can invert the function $h=h(p)$ and find $p$ as a function of $h$. Then using the result \eqref{hhpp} one finds $p$ as a function of $r$ and obtains the metric function $f=f(r)$
\be
f=1-r^2 p\, .
\ee

Let us summarize. The reduced action for quasi-topological gravity coupled to nonlinear electrodynamics is specified by two functions, $h = h(p)$ and $\mathcal{L} = \mathcal{L}(\mathcal{E})$, which parametrize the theory. A solution $(f,\varphi,N)$ of the resulting system of ordinary differential equations can be expressed in terms of integrals. To obtain an explicit form of the solution, however, one must also solve the algebraic problem of inverting the functions $h(p)$ and $\mathcal{L}(\mathcal{E})$.
In what follows, we consider the case of Born–Infeld electrodynamics coupled to quasi-topological gravity. We keep the number of spacetime dimensions arbitrary, with $D \geq 4$.
\subsection{Reissner-Nordstr\"om-Tangherlini solution}\label{Sec_RNT}

In the previous subsection, we derived the field equations for a coupled system of quasi-topological gravity (QTG) and nonlinear electrodynamics using the spherically reduced action.
Before discussing solutions of the general system of equations \eqref{SEQ}, we briefly review the well-known solutions describing higher-dimensional static charged black holes in the Einstein–Maxwell theory, which are described by
\be
h(p)=p\hh \CAL{L}=\dfrac{1}{2}\CAL{E}^2\, .
\ee

In the absence of an electric field, one has
\be
h(p)=\dfrac{\mu}{r^{D-1}}\, .
\ee
For the Einstein equations $h(p)=p$ and the metric function is
\be
f=1-\Big(\dfrac{r_g}{r}\Big)^{D-3}\, .
\ee
The corresponding metric is the Tangherlini-Schwarzschild solution describing a static spherically symmetric black hole with gravitational radius $r=r_g=\mu^{1/(D-3)}$.

For a non-vanishing electric charge $Q$
\be
\CAL{E}=\dfrac{Q}{r^{D-2}} \, ,
\ee
and the function $U(r)$, which enters \eqref{rrff} is
\be
U= - \dfrac{\varkappa Q^2}{(D-2) r^{D-2}}\, .
\ee
This gives
\be
Z= -\dfrac{\varkappa Q^2}{(D-2)(D-3)}\dfrac{1}{r^{D-3}}\, .
\ee
The metric function $f$ for this solution is
\be \n{MAXW}
f=1-\dfrac{\mu}{r^{D-3}}+\dfrac{\varkappa Q^2}{(D-2)(D-3)}\dfrac{1}{r^{2(D-3)}}\, .
\ee
This metric function gives the Reissner-Nordstr\"om-Tangherlini metric which is a solution for a higher dimensional static charged black hole of the Einstein-Maxwell theory.
Quite often one introduces the quantity $\hat{Q}$ related to $Q$ as follows (see e.g. \cite{Kodama:2003kk})
\be
\hat{Q}^2=\dfrac{\varkappa Q^2}{(D-2)(D-3)}\, ,
\ee
so that \eqref{MAXW} takes the form
\be
f=1-\dfrac{\mu}{r^{D-3}}+\dfrac{\hat{Q}^2}{r^{2(D-3)}}\, .
\ee
One also has
\be\n{pprr}
p=\dfrac{1}{r^{D-1}}\Big(
\mu -\dfrac{\hat{Q}^2}{r^{D-3}}
\Big)\, .
\ee
The variable $\hat{Q}$ has the same dimensions as $\mu$, that is $[\hat{Q}]=Length^{D-3}$, while the ratio $\hat{Q}/\mu$ is dimensionless.

For $\hat{Q}\le \mu/2$ the metric \eqref{MAXW} describes a charged static black hole. Its horizons are located at $r=r_{\pm}$, where
\be
r_{\pm}^{D-3}=\dfrac{\mu}{2}\pm \sqrt{(\mu/2)^2-\hat{Q}^2}\, .
\ee
The metric function \eqref{MAXW} can be written in the form
\be
f=1-\dfrac{m(r)}{r^{D-3}}\,
\ee
where $m(r)$ is a so called mass function
\be
m(r)=\mu-\dfrac{\hat{Q}^2}{r^{D-3}}\,.
\ee
This function is directly related to the higher dimensional Misner-Sharp mass $M_{MS}$ \footnote{
The Misner–Sharp mass is a quasi-local notion of gravitational energy defined in spherically
symmetric spacetimes.
In four dimensions it was introduced by Misner and Sharp in their
study of relativistic gravitational collapse \cite{Misner-Sharp} .
The construction admits a natural extension to
higher-dimensional Einstein gravity and also to Lovelock, Gauss–Bonnet, and quasitopological
theories \cite{ Hayward1996,Hayward1998,MaedaNozawa2008a,MaedaNozawa2008b,EmparanReall2008}.} 
\be
m(r)=\dfrac{16\pi G^D}{(D-2)\Omega_{D-2}} M_{MS}\, .
\ee
For $r=r_0$, which satisfies the relation
\be
r_0^{D-3}=\dfrac{\hat{Q}^2}{\mu} \, ,
\ee
the mass function, and hence the Misner-Sharp mass, vanishes. This property is commonly interpreted as stating that the total energy of the electric field located outside $r=r_0$ is equal to the mass parameter $\mu$ measured at infinity.

Let us emphasize that there exists a simple relation between the mass function $m$ and the primary curvature invariant $p$
\be
p=\dfrac{m}{r^{D-1}}\, .
\ee
This implies that when the mass function $m$ vanishes, the curvature invariant $p$ vanishes as well. Let us emphasize that, since the mass function differs from the Misner--Sharp mass only by a constant factor, the primary curvature invariant p vanishes on the sphere where the Misner--Sharp mass vanishes. This observation provides a more geometric interpretation of the surface $r = r_0$. To the best of our knowledge, this surface does not have a standard name in the literature. Since it plays an important role in what follows, we introduce the term \emph{$p$-sphere} to denote this surface and its analogue in QTG models.
Such QTG models are specified by a function $h=h(p)$, which for small $p$ behaves as $h(p)=p+O(p^2)$. This property ensures the correct Einstein gravity limit in the weak-field regime. It also implies that the condition $h(p)=0$ can be used to define the $p$-sphere in QTG.

\section{Born-Infeld electromagnetic field}\label{sec3}

\subsection{The model and its solution}

To describe nonlinear electrodynamics, we adopt the well-known and widely studied Born–Infeld model. In this case, the spherically reduced Lagrangian density $\mathcal{L}$ takes the form
\begin{equation}
\CAL{L}(\mathcal{E}) = b^2\left(1 - \sqrt{1 - \frac{\mathcal{E}^2}{b^2}} \right)\, .
\end{equation}
As early,  $\mathcal{F} = \CAL{E}^2$.
For large $b$ one has
\be
\CAL{L}(\mathcal{E})=\dfrac{1}{2}\mathcal{E}^2\, ,
\ee
and this model correctly reproduces Maxwell's equations in the weak field regime. For a point-like charge, the solution of this model remains finite and the value of the electric field $\CAL{E}$ is (uniformly) bounded at the position of the charge by the value $b$. This means that
the parameter $b$ has the meaning of the maximal value of the electric field.

Taking the derivative of $\CAL{L}$ over $\mathcal{E}$, one finds
\begin{equation}
\frac{d\CAL{L}}{d\mathcal{E}} = \frac{\CAL{E}}{\sqrt{ 1 - \frac{\CAL{E}^2}{b^2} }}\ \, .
\end{equation}
Substituting this into \eqref{eq_ED3} allows us to solve the obtained equation for $\CAL{E}$ and obtain the following expression
\begin{equation}\label{eq_PhiBorn}
\CAL{E} = \frac{Q}{\sqrt{r^{2D-4} + \frac{Q^2}{b^2}}} \, .
\end{equation}
The function $\CAL{E}$ has the following asymptotics
\begin{itemize}
\item For $r \to \infty$, \quad
$\CAL{E}= \dfrac{Q}{r^{D-2}} + O(1/r^{3(D-2)}) $;
\item   For $r \to 0$, \quad $\CAL{E} = b + O(r^{2(D-2)})$ \, .
\end{itemize}
The latter expression implies
that as $r$ approaches the position of the source, the function $\CAL{E}$ will reach the maximum value of the electric field $b$.

Using \eqref{eq_PhiBorn}, one can write the Lagrangian density  $\CAL{L}$ as follows
\begin{equation}\label{eq_PBorn}
\CAL{L} = b^2 \left(1-\frac{r^{D-2}}{\sqrt{r^{2D-4} + Q^2/b^2}}\right) \, .
\end{equation}
Substituting \eqref{eq_PhiBorn} and \eqref{eq_PBorn} into \eqref{rrff} one finds the function $U=U(r)$
\be\n{eqU}
U(r)= -\frac{2\varkappa b^2}{D-2} \bigg( \sqrt{r^{2D-4} + Q^2/b^2} - r^{D-2} \bigg) \, .
\ee

\subsection{Dimensionless form of the equations}

It is now convenient to write the main equations \eqref{hhpp}, \eqref{eq_PhiBorn} and \eqref{eqU} in dimensionless form. For this purpose, we make the following change of variables
\be\n{CHANGE}
Q=b q
\hh r=(Q\rho/b)^{1/(D-2)} .
\ee
Then the expression \eqref{eq_PhiBorn} for the electric field $\CAL{E}$ takes the form
\be \n{CE}
\CAL{E}=\dfrac{b}{\sqrt{1+\rho^2}}\, .
\ee
Let us emphasize that this simple expression follows from the special scaling properties of the Born–Infeld Lagrangian density. Indeed, one expects the Coulomb field $\mathcal{E}$ to depend on three variables, $b$, $Q$, and $r$. By the $\Pi$-theorem, it can be written in the form $\mathcal{E}=b g$, where $g$ is a function of two independent dimensionless combinations constructed from $b$, $Q$, and $r$. As we have seen, however, $g$ in fact depends only on a single dimensionless parameter, $\rho$.
Similarly, using \eqref{CE}, one obtains
\be \n{FFFF}
\begin{split}
\CAL{L}&=b^2 \Big[
1-\dfrac{\rho}{\sqrt{\rho^2+1}}
\Big]\, ,\\
U&= -\frac{2\varkappa b^2 q}{D-2}\Big( \sqrt{\rho^2+1}-\rho\Big)\, .
\end{split}
\ee

\begin{figure}[!hbt]
    \centering
    \includegraphics[width = 0.8 \linewidth]{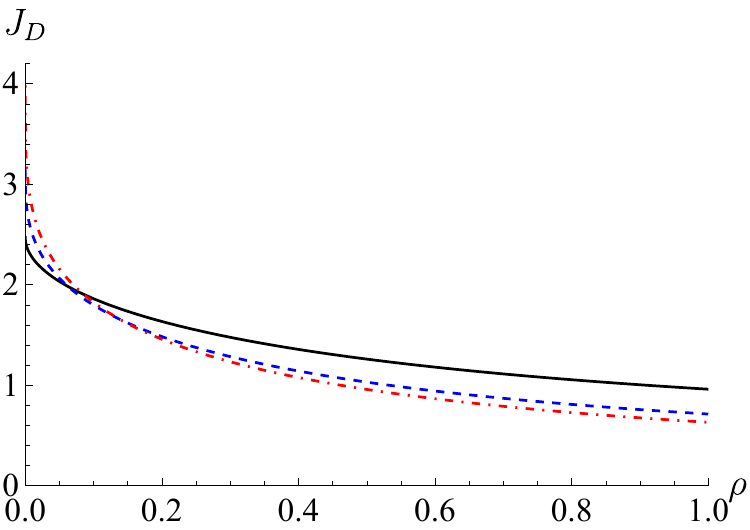}
    \caption{{
    Plot of the function $J_D(\rho)$  for $D=4$ (solid), $D=5$ (dashed) and $D=6$ (dot-dashed) } }
    \label{FIG_JD}
\end{figure}

Using the relation
\be
dr=\dfrac{1}{D-2} \Big( \dfrac{Q}{b}\Big)^{1/(D-2)} \rho^{-(D-3)/(D-2)} d\rho
\ee
one gets

\be\n{JJJDD}
\begin{split}
&Z= -\dfrac{2\varkappa b^2q^{(D-1)/(D-2)}}{(D-2)^2} J_D(\rho)\, ,\\
&J_D(\rho) =\int_{\rho}^{\infty} dx x^{(3-D)/(D-2)} \Big( \sqrt{x^2 +1}-x\Big)\, .
\end{split}
\ee
Figure~\ref{FIG_JD} shows plots of the function $J_D(\rho)$ for spacetime dimensions $D=4$, $D=5$ and $D=6$.

The integral representation for $J_D(\rho)$ in \eqref{JJJDD} implies that it is a positive function of $\rho$ which monotonically decreases from $J_D^{(0)}=J_D(\rho=0)$ at $\rho=0$ until it reaches 0 at $\rho\to \infty$.

The asymptotic form of $J_D(\rho)$ for large $\rho$ is
\be\n{BIG}
J_D(\rho) \approx  \frac{D-2}{2(D-3)}\rho^{(3-D)/(D-2)}\, .
\ee
It can be easily found by
calculating the integral in \eqref{JJJDD} using the asymptotic
\be
\sqrt{x^2+1}-x\approx \dfrac{1}{2x}\, .
\ee

For $\rho=0$ the integral for $J_D$ in \eqref{JJJDD} can be expressed in terms of $\Gamma$-functions
\begin{equation}\n{JJ00}
\begin{split}
J_D^{(0)}& =J_D(\rho=0)\\
&=\frac{D-2}{2(D-1)\sqrt{\pi}}\,
\Gamma\!\left(\frac{1}{2(D-2)}\right)
\Gamma\!\left(\frac{D-3}{2(D-2)}\right) \, ,\\
&=\quad \frac{\pi (D-2)^2}{(D-1)\,2^{\frac{D-3}{D-2}}
\cos\!\left(\frac{\pi}{2(D-2)}\right)\,
\Gamma\!\left(\frac{1}{D-2}\right)
} \, .
\end{split}
\end{equation}
Some of the values of the function $J_D^{(0)}$ are
\be
\begin{split}
J_4^{(0)}&=\frac{\Gamma\!\left(\tfrac14\right)^2}{3\sqrt{\pi}}\approx 2.47210\, ,\\
J_5^{(0)}&=\frac{3}{8\sqrt{\pi}}\,
\Gamma\!\left(\tfrac16\right)\Gamma\!\left(\tfrac13\right)\approx 3.15491\, ,\\
J_6^{(0)}&=
\frac{2}{5\sqrt{\pi}}\,
\Gamma\!\left(\tfrac18\right)\Gamma\!\left(\tfrac38\right)
\approx4.03028 \, .
\end{split}
\ee

For an arbitrary $\rho$, the integral $J_D(\rho)$ can also be calculated analytically. One has
\be \n{JJDD}
\begin{split}
J_D(\rho)&=
\frac{D-2}{D-1}\,\rho^{\frac{D-1}{D-2}}\Big( 1-F_D(\rho)\Big)\, ,\\
F_D(z)&={}_2F_{1}\!\Big(
-\frac{1}{2},\,-\frac{D-1}{2(D-2)};
\frac{D-3}{2(D-2)};\,-1/z^2
\Big)\, .
\end{split}
\ee
Here ${}_2F_1$ denotes the Gauss hypergeometric function \cite{NIST2010}.
The hypergeometric function
${}_2F_{1}\!(a,b;c;z)$ has the following asymptotic form at $z=0$
\be
{}_2F_{1}(a,b;c;z)=1+\dfrac{a b}{c} z+\ldots \, .
\ee

Let us define a suitable dimensionless form of the function $h(p)$ and $p$
\be
\hat{h}=\hat{h}(\hat{p})\hh \hat{p}=\ell^2 p\, .
\ee
Using \eqref{JJJDD} and \eqref{JJDD} one can write the solution for $\hat{h}$
\be \n{hhhh}
\begin{split}
\hat{h}&= \rho^{-(D-1)/(D-2)}\Big[
A-\dfrac{D-1}{D-2}\beta J_D(\rho)\Big]  \\
&=A\rho^{-(D-1)/(D-2)}+\beta(F_D(\rho)-1)\, ,\\
A&=\dfrac{\ell^2\mu}{q^{(D-1)/(D-2)}}\hh
\beta =\dfrac{2\varkappa b^2\ell^2}{(D-1)(D-2)}\, .
\end{split}
\ee

The function $\hat{h}=\hat{h}(\hat{p})$ specifies the concrete version of the QTG model. This relation can be used to get $\hat{p}=\hat{p}(\hat{h})$. The function $f$ which enters the metric can be written in the following dimensionless form
\be \n{ffqq}
f=1-\hat{q}^2 \rho^{2/(D-2)} \hat{p}(\rho)\hh
\hat{q}=\dfrac{q^{1/(D-2)}}{\ell}\, .
\ee
The dimensionless parameter $A$ which enters the expression \eqref{hhhh} for $\hat{h}$ can be written as follows
\be
A=\dfrac{\mu}{\ell^{D-3}}\dfrac{1}{\hat{q}^{D-1}}\, .
\ee

Let us note that the metric function contains 3 dimensionless parameters. One of them, $\beta$, is determined through the action of the theory, while the other two, $A$ and $\hat{q}$, specify a solution. These two parameters are related to the mass parameter $\mu$ of the black hole and its charge $\hat{Q}$. For the parameter $\mu$ one has
\be
\mu=\ell^{D-3}\hat{q}^{D-1}A\, .
\ee
We denote a dimensionless version of $\mu$ by $\hat{\mu}$
\be
\hat{\mu}=\dfrac{\mu}{\ell^{D-3}}=\hat{q}^{D-1}A\, .
\ee
We also denote by $\sigma$ the charge-to-mass ratio
\be
\sigma=\dfrac{2\hat{Q}}{\mu}\, .
\ee
It is defined such that, for the Tangherlini–Reissner–Nordström metric \eqref{MAXW}, the condition for an extremally charged black hole corresponds to $\sigma = 1$.

\subsection{Properties of the solution: Universality and $p$-spheres.}

The remarkable feature of the solution \eqref{hhhh} is its universality. In particular, it is independent of the specific choice of the QTG model and is uniquely determined by two dimensionless parameters, $\beta$ and $A$. This universality enables a detailed and systematic analysis of its properties.
For this analysis, it is convenient to rewrite $\hat{h}$ in the form
\be \n{hhhh0}
\begin{split}
\hat{h}&= \dfrac{D-1}{D-2}\beta \rho^{-(D-1)/(D-2)}\Big[ \hat{A}-J_D(\rho)\Big]\, ,\\
\hat{A}&=\dfrac{(D-2)A}{(D-1)\beta}\, .
\end{split}
\ee
The expression inside the square brackets is a monotonically increasing function of $\rho$, varying from $\hat{A}-J_D^{(0)}$ at $\rho=0$ to $\hat{A}$ as $\rho \to \infty$. Since the parameter $\hat{A}$ is positive, $\hat{h}(\rho)$ is positive at large $\rho$, where $J_D(\rho)$ becomes small.

If $\hat{A}-J_D^{(0)}>0$, this function, and hence $\hat{h}$, remain positive over the entire semi-axis $\rho \ge 0$. In this case, $\hat{h}$ diverges as $\rho \to 0$.

\begin{figure}[!hbt]
    \centering
    \includegraphics[width=0.8
\linewidth]{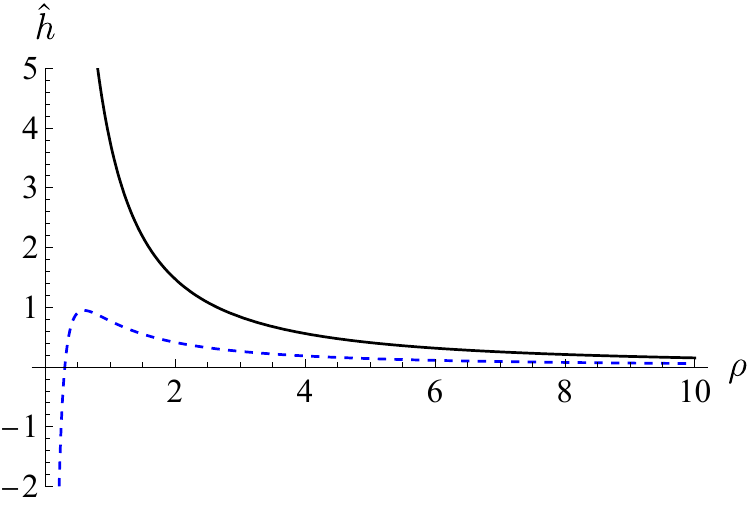}
    \caption{Plot of the function $\hat{h}$ from \eqref{hhhh0} for
    $\hat{A}-J_4^{(0)} = 1$ (solid) and for $\hat{A}-J_4^{(0)} = -1$ (dashed). One can see that the plot becomes negative at some point $\rho = \rho_0$ when $\hat{A}-J_D^{(0)} < 0$, indicating the existence of a $p$-sphere. }
    \label{FIG_h_pm}
\end{figure}

In the opposite case, $\hat{A}-J_D^{(0)}<0$, the function $\hat{h}$ crosses zero at some $\rho=\rho_0$ and becomes negative for $\rho<\rho_0$. This indicates that the solution possesses a $p$-sphere at $\rho=\rho_0$ (see the discussion at the end of subsection~\ref{Sec_RNT}). Inside the $p$-sphere, $\hat{h}$ remains negative and diverges as $\rho \to 0$. Figure~\ref{FIG_h_pm} illustrates the behavior of $\hat{h}(\rho)$ in these two cases.

A critical condition separating these two different regimes is the following
\be \n{hhAA}
\hat{A}=J_D^{(0)}\, ,
\ee
or, equivalently,
\be
A=\dfrac{D-1}{D-2}\beta J_D^{(0)}\, .
\ee
Using the expressions for $A$, $\hat{\mu}$ and $\sigma$ one obtains the following equation for $A$
\begin{equation}\n{AAA}
    A = \dfrac{1}{(\hat{\mu}
    \sigma^{D-1})^{1/(D-2)} }
    \bigg(\frac{(D-3)}{2\beta(D-1)} \bigg)^{(1-D)/(2(D-2))}\, .
\end{equation}
Using this expression, the condition for the existence of a $p$-sphere, and consequently for the presence of a curvature singularity, can be written in the form
\be \n{SING}
\begin{split}
&\hat{\mu}    \sigma^{D-1}>\CAL{C}  \, ,\\
\CAL{C}=\dfrac{2^{(D-1)/2}}{\beta^{(D-3)/2} (J_D^{(0)})^{D-2}}&\dfrac{(D-2)^{D-2}}{(D-1)^{(D-3)/2}(D-3)^{(D-1)/2}}
\end{split}
\ee
Some of the values of the constant $\CAL{C}$ for different dimensions are
\begin{equation}
\begin{split}
    \CAL{C}|_{D=4} &= \frac{24\sqrt{6} \pi}{\beta^{1/2} \Gamma(1/4)^4} \sim  1.06884 \frac{1}{\beta^{1/2}}\, , \\
    \CAL{C}|_{D=5} &= \frac{128 \pi^{3/2}}{\beta \Gamma(1/6)^3 \Gamma(1/3)^3} \sim 0.21495 \frac{1}{\beta} \, , \\
    \CAL{C}|_{D=6} &= \frac{1600 \sqrt{10/3}\pi^2}{9\beta^{3/2}\Gamma(1/8)^4 \Gamma(3/8)^4} \sim 0.03149 \frac{1}{\beta^{3/2}} \, .
\end{split}
\end{equation}

Let us emphasize that the expression \eqref{AAA} for $A$ and the condition \eqref{SING} for the existence of a $p$-sphere do not depend on a specific choice of the QTG model and are, in this sense, universal.
In the next section, we show that in QTG models where uncharged black holes are regular, their charged counterparts can be either regular or singular. In particular, for Hayward-type QTG black holes, the existence of a $p$-sphere implies the presence of a curvature singularity in the interior. By contrast, in Born–Infeld QTG, charged black holes that possess a $p$-sphere remain regular.

For $\beta$ of order unity, the parameter $\mathcal{C}$ entering this relation is likewise of order unity in $D=4$.

\section{Black holes in QTG with a nonlinear electromagnetic field.}\label{sec4}

\subsection{Regular vs singular black holes}

A QTG model is specified by choosing a function $h=h(p)$. For Einstein gravity $h(p)=p$. For QTG one deals with $h(p)$ presented in the form
\be\label{h_def}
h(p)=p+\sum_{j=2}^{\infty}\alpha_j \ell^{2(j-1)}p^j\, .
\ee
In \cite{QT_BH}, it was shown that regular vacuum black holes solutions can exist if the series is not terminated at finite $j$ and special conditions are imposed on the coefficients $\alpha_j$.

Let us remind that the parameter $p$ used in this paper denotes the primary basic curvature invariant. In fact, a static spherically symmetric spacetime with the metric \eqref{MMM} has 4 algebraically independent scalar curvature invariants (see e.g. \cite{rbh_pfkz}). They can be easily identified by writing the independent components of the Riemann tensor. We denote these 4 basic curvatures invariants by $p$, $q$, $u$ and $v$
\be
p=\dfrac{1-f}{r^2}\hhh
q=\dfrac{{}^{(2)}\Box r}{r}\hhh
 u=\dfrac{r^{,\mu}r^{,\nu} r_{;\mu\nu}}{r}\hhh v=R^{(2)}\, .
\ee
Here ${}^{(2)}\Box$ and $R^{(2)}$ are the box operator and curvature in the $(t,r)$-slice of the metric. For $N=1$, one has that $q=u$ and this number reduces to 3. For the metric \eqref{MMM} with $N=1$, there exist the following differential relations between the basic curvature invariants
\be\label{invars1}
\begin{split}
q&=u=p+\frac{1}{2}r\partial_r p\, ,\\
v&=p+2r\partial_r p+\frac{1}{2}r^2\partial^2_r p\, .
\end{split}
\ee
For $N=1$, the basic curvature invariants are constructed from a single metric function and its derivatives; the only difference lies in the number of such derivatives.
The relations \eqref{invars1} imply that if the primary curvature invariant $p$ is regular, bounded, and smooth on the interval $\rho\in [0,\infty)$, so are the other basic curvature invariants $q$, $u$ and $v$. Thus, any polynomial scalar invariants constructed from the Riemann curvature tensor are also regular and uniformly bounded.
In particular, suppose the function $p=p(r)$ is finite and regular at $r=0$ and has the following expansion, with positive integer powers of $r$
\be
p(r)=p_0+p_1 r+\dfrac{1}{2}p_2 r^2+\ldots\, .
\ee
Then one has
\be\label{invrho}
\begin{split}
q&=u=p_0+\dfrac{3}{2}p_1 r+p_2 r^2 \, ,\\
v&=p_0+ 3p_1 r+ 3 p_2 r^2\, .
\end{split}
\ee

The expressions \eqref{invars1} can be also written in the dimensionless coordinate $\rho$
\be
\begin{split}
q&=u=p+\frac{D-2}{2}\rho\pa_{\rho} p\, ,\\
v&=p+2(D-2)\rho\pa_{\rho} p+ \frac{1}{2}(D-2)^2  \rho^2\pa^2_{\rho} p\, .
\end{split}
\ee 
After these general remarks, let us return to our problem. As we already mentioned, it is convenient to use the dimensionless form of $h$ and $p$
\be
\hat{h}=\ell^2 h\hh \hat{p}=\ell^2 p\, .
\ee
After obtaining the function $\hat{h}(\rho)$ given by \eqref{hhhh}, one still needs to determine the corresponding dimensionless primary curvature invariant $\hat{p}$. This requires inverting the relation $\hat{h}=\hat{h}(\hat{p})$ to obtain $\hat{p}=\hat{p}(\hat{h})$. Substituting $\hat{h}(\rho)$ then yields $\hat{p}(\rho)$, which in turn allows one to reconstruct the metric function via \eqref{ffqq}.

This reconstruction crucially depends on the existence of a $p$-sphere. As shown above, in the absence of a $p$-sphere the solution \eqref{hhhh} for $\hat{h}$ is positive for $\rho \in [0,\infty)$, and regularity requires only that the relation $\hat{h}=\hat{h}(\hat{p})$ admits a regular inverse for $\hat{h}>0$.

In the presence of a $p$-sphere, however, the situation is qualitatively different. For the metric to remain regular, the solution $\hat{h}(\rho)$ must admit a regular inverse over the entire axis $\rho \in (-\infty,\infty)$.

To illustrate this point, in what follows we consider two special versions of QTG models, often used in the discussion of regular black holes:
\begin{itemize}
\item The Hayward-type QTG model is specified by setting $\alpha_j = 1$  in \eqref{h_def}. This choice specifies the function $\hat{h}(\hat{p})$, which in turn determines the basic curvature invariant $\hat{p}$ characterizing the model:
\be \label{QHH}
\hat{h}=\dfrac{\hat{p}}{1-\hat{p}}\quad \to \quad \hat{p}=\dfrac{\hat{h}}{1+\hat{h}} \, .
\ee
\item The Born-Infeld-type QTG model, defined by now using
\begin{equation}
\alpha_j = \frac{(1-(-1)^j)\Gamma(j/2)}{2\sqrt{\pi}\Gamma((1+j)/2)} \, ,
\end{equation}
in equation \eqref{h_def}. One gets
\be \label{QBI}
\hat{h}=\dfrac{\hat{p}}{\sqrt{1-\hat{p}^2}}\quad \to \quad
\hat{p}=\dfrac{\hat{h}}{\sqrt{1+\hat{h}^2}} \, .
\ee
\end{itemize}
For both models, the vacuum black hole solutions are regular. However, as we will show, the properties of charged black holes are markedly different.

\subsection{Hayward-type charged black holes}

\begin{figure}[!hbt]
    \centering
    \includegraphics[width=0.8
\linewidth]{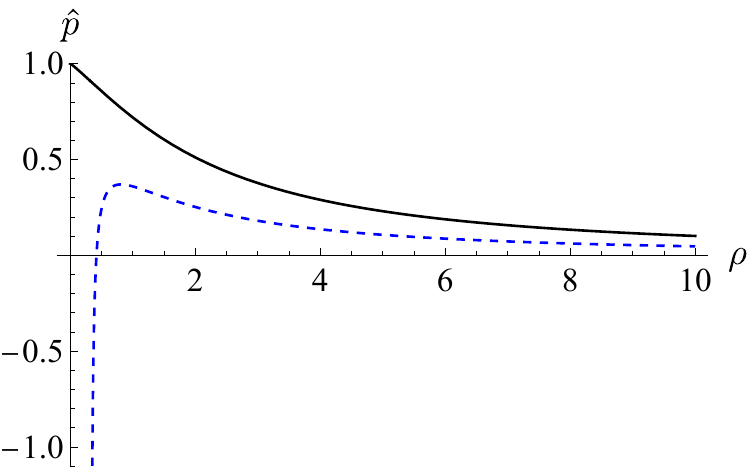}
    \caption{Plot of the primary curvature invariant $\hat{p}$ for the Hayward-type QTG model  as a function of $\rho$ for $\beta=1$ and $D=4$. Two cases are shown: $A=2$ (dashed) and $A=4$ (solid). In the latter case $\hat{p}$ diverges when $A<\tfrac{3}{2}\beta J_4^{(0)}$.}
    \label{FIG_invpH}
\end{figure}

Consider the solution \eqref{hhhh} of Hayward-type QTG, which has a $p$-sphere. As we discussed earlier, this is a generic case when the mass of the black hole is not miscroscopically small and the electric change is not extremely small, so that the inequality \eqref{SING} is valid. Inside the $p$-sphere, $\hat{h}$ is negative and diverges as $\rho \to 0$. Hence, there exists a radius $\rho=\rho_0^*$ at which $\hat{h}$ attains the value $-1$.
Using \eqref{QHH}, one concludes that the primary curvature invariant (as well as other basic curvature invariants) diverges at this point.
Figure \ref{FIG_invpH} displays this principal curvature invariant for different values of $A$.

In other words, within this QTG model, the standard curvature singularity at $\rho=0$ is effectively “shifted” to a sphere of finite radius. This behaviour contrasts sharply with the neutral regular black hole case, where the solutions remain nonsingular.
We note, however, that regular charged black hole solutions in the Hayward-type QTG model still exist when the inequality \eqref{SING} is violated.

\begin{figure}[!hbt]
    \centering
    \includegraphics[width=0.8
\linewidth]{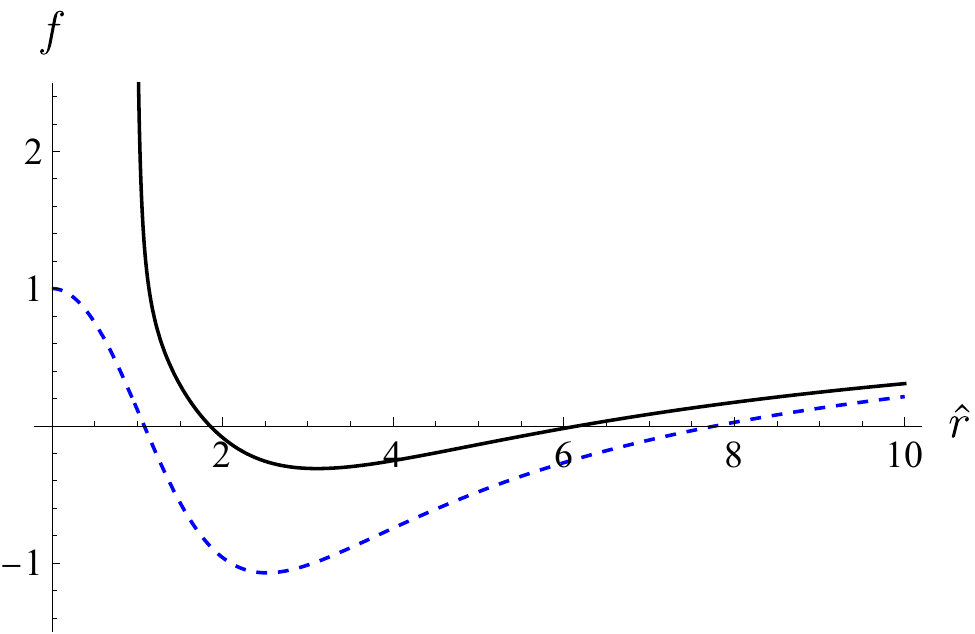}
    \caption{Plot of the metric function $f(\hat{r})$, $\hat{r}=r/\ell$, for the Hayward-type QTG model for $\beta = 1$ and $D=4$. Two cases are displayed for different $(\hat{\mu},\sigma)$ parameters, a divergent metric with parameters $(8,0.8)$ (solid) and a regular black hole with parameters $(8,0.2)$ (dashed)}\label{FIG_metricH}
\end{figure}

Figure~\ref{FIG_metricH} shows the metric function $f$ as a function of the dimensionless coordinate
$\hat{r} = r/\ell$ for singular and regular charged black holes in the Hayward-type QTG. Similar results, as well as their thermodynamic properties, can be found in \cite{Hennigar2025}.

\subsection{Born-Infeld-type charged black holes}

\begin{figure}[!hbt]
    \centering
    \includegraphics[width=0.8
\linewidth]{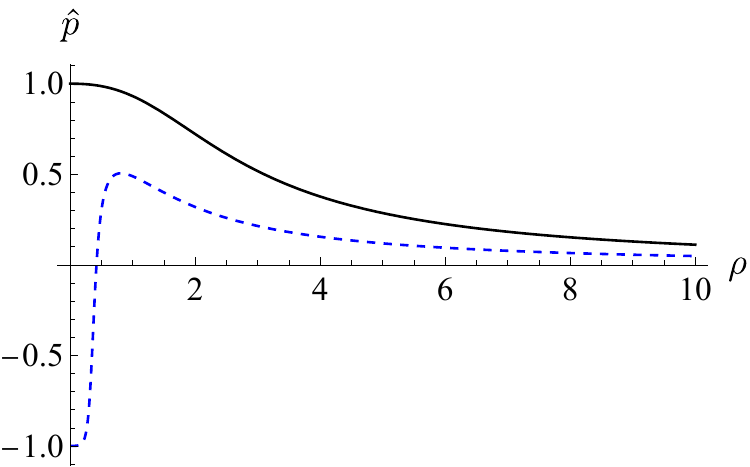}
    \caption{Plot of the curvature invariant $\hat{p}$ in the Born–Infeld–type QTG model as a function of $\rho$ for $\beta=1$ and $D=4$. Two cases are shown: $A=2$ (dashed) and $A=4$ (solid). In the latter  $\hat{p}$ becomes negative at the origin, where it approaches the finite value of $-1$.}
    \label{FIG_invpBI}
\end{figure}

In the Born–Infeld–type QTG model, the situation is qualitatively different. Equation \eqref{QBI} shows that, for all values $\hat{h} \in (-\infty,\infty)$, the primary curvature invariant $\hat{p}$ remains a well-defined, finite, and regular function of $\rho$. The function $\hat{p}=\hat{p}(\rho)$ simply changes sign at the $p$-sphere and remains negative inside it. In particular, at $\rho=0$, where $\hat{h}$ is negative and divergent, $\hat{p}(\rho)$ approaches the finite value of $-1$. Figure~\ref{FIG_invpBI} illustrates the behaviour of the invariant $\hat{p}$ for different values of the parameter $A$.

\begin{figure}[!hbt]
    \centering
    \includegraphics[width=0.8
\linewidth]{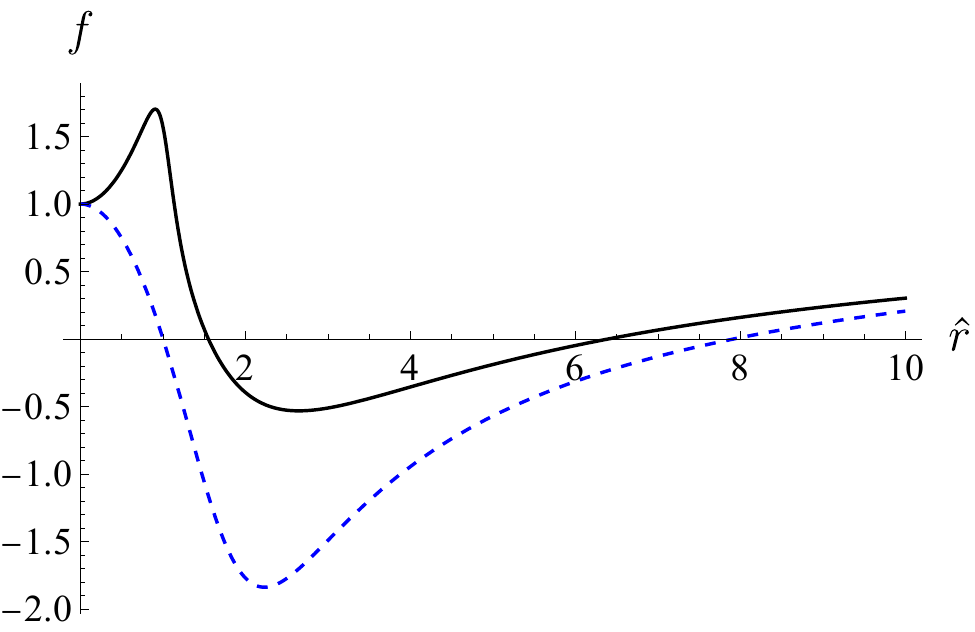}
    \caption{Plot of the metric function $f(\hat{r})$, with $ \hat{r}=r/\ell $, for the Born-Infeld-type QTG model with $ \beta=1 $ in $ D=4 $. Two cases are shown for different values of $ (\hat{\mu},\sigma) $: the solid curve corresponds to $ (\hat{\mu},\sigma)=(8,0.8) $, yielding an anti-de Sitter core, while the dashed curve corresponds to $ (\hat{\mu},\sigma)=(8,0.2) $, which yields a de Sitter core.
}\label{FIG_metricQBI}
\end{figure}

\begin{figure}[!hbt]
    \centering
    \includegraphics[width=0.8
\linewidth]{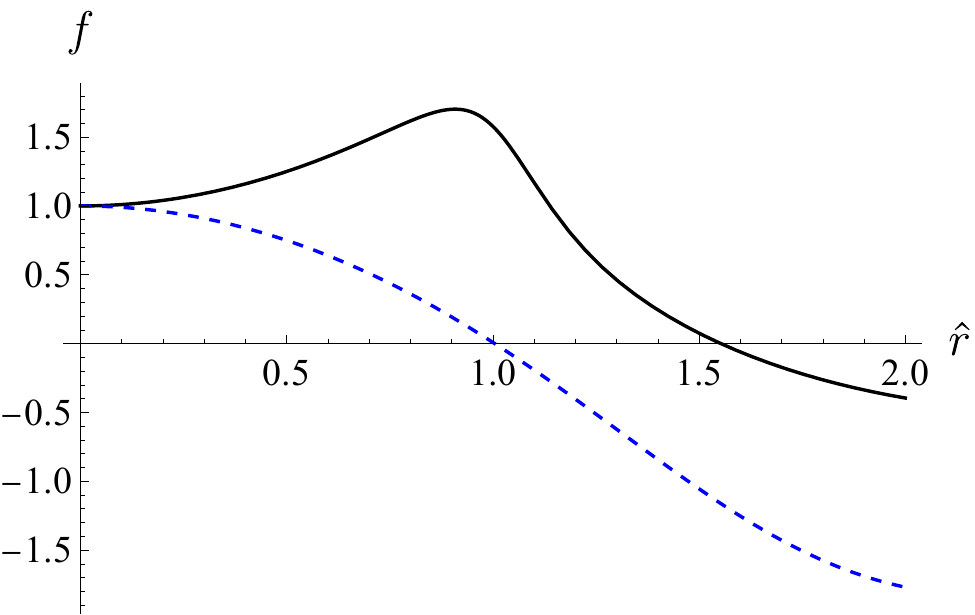}
    \caption{The behaviour of the metric function $ f(\hat{r}) $ near $ \hat{r}=0 $ (with $ \hat{r}=r/\ell $) is shown for the Born-Infeld-type QTG model. The solid curve corresponds to the parameters $ (\hat{\mu},\sigma) = (8,0.8)$ associated with an anti-de Sitter core, while the dashed curve corresponds to $ (\hat{\mu},\sigma)=(8,0.2)$, for which the core is de Sitter. }\label{FIG_metricQBI_r0}
\end{figure}

The asymptotic form of the metric function $f(r)$, written near the center at $r=0$ is
\be
\n{assf}
f(r)=1+\dfrac{r^2}{\ell^2}+\ldots \, .
\ee
The metric is regular, and the asymptotic behavior \eqref{assf} is universal. It follows that charged black hole solutions in the Born–Infeld–type QTG model are always regular.
Interestingly, in contrast to neutral regular black holes, which typically possess a de Sitter core, regular charged black holes exhibit an anti–de Sitter core. However, when the inequality \eqref{SING} is violated, the solution retains a de Sitter core. This behaviour is illustrated on Figures~\ref{FIG_metricQBI} and \ref{FIG_metricQBI_r0} for the metric function $f$ as a function of $\hat{r}$.

\section{Extremely charged black holes}\label{sec_extremal}

\subsection{Extremality condition}

In General Relativity, charged black holes exist only if their charge $\hat{Q}$
satisfies the condition
\be
\sigma=2\hat{Q}/\mu \le 1\,
\ee
where $\mu$ is the black hole mass parameter.If this condition is violated, the corresponding solution describes a naked singularity. The charged solutions in QTG models can likewise represent either black holes or naked singularities. However, the extremality condition $\sigma=1$ is modified. In this section, we discuss the corresponding extremality condition.
This condition is defined by the following equations
\begin{equation}\label{critical_conds}
    f|_{\rho_*} = 0\, , \quad \frac{df}{d\rho}\bigg|_{\rho_*} = 0 \, .
\end{equation}
Here $f$ is the metric function. For a regular black hole, the first equation determines the locations of the outer and inner horizons, while the second applies when these two horizons coincide. In this case, the surface gravity of the black hole vanishes. We refer to Eqs.~\eqref{critical_conds} as the extremality conditions.

As earlier, we write the metric function $f$ the  form
\begin{equation}\label{fff}
  f=  1 -\hat{q}^2\rho^{2/(D-2)} \hat{p}(\hat{h}) \, .
\end{equation}
Differenciating $f$ with respect to $\rho$ one gets
\begin{equation}\label{eq_df}
    \frac{df}{d\rho} = -\hat{q}^2\rho^{2/D-2}
    \Big[
    \dfrac{2}{D-2}\dfrac{\hat{p}}{\rho}+\dfrac{d\hat{h}}{d\rho}\dfrac{d\hat{p}}{d\hat{h}}
    \Big]\, .
\end{equation}
Then, the second condition in \eqref{critical_conds} can be written as follows
\be \n{cond2}
 \dfrac{2}{D-2}\dfrac{\hat{p}}{\rho}+\dfrac{d\hat{h}}{d\rho}\dfrac{d\hat{p}}{d\hat{h}}=0\, .
\ee
To calculate $d\hat{h}/d\rho$, which enters this relation, it is convenient to write $\hat{h}$ in the form
\be \n{hhhhh}
\begin{split}
&\hat{h}=\rho^{(1-D)/(D-2)}\Big[
A - \beta \dfrac{D-1}{D-2}J_D(\rho) \Big] \, , \\
& \beta = \frac{2\varkappa b^2\ell^2}{(D-1)(D-2)} \, .
\end{split}
\ee
and use the definition of $J_D(\rho)$ given in \eqref{JJJDD} which implies
\be
\dfrac{dJ_D}{d\rho}= -\rho^{(3-D)/(D-2)}(\sqrt{\rho^2+1}-\rho)\, .
\ee
Simple calculations give the following result
\be
\dfrac{d\hat{h}}{d\rho}=\dfrac{D-1}{D-2}\dfrac{1}{\rho^2}\Big[ -\rho\hat{h}+\beta (\sqrt{\rho^2+1}-\rho)\Big]\, .
\ee
Using this result, one can write the condition \eqref{cond2} in the form
\be \n{second}
2\rho \hat{p} +(D-1)\Big[ -\rho\hat{h}+\beta (\sqrt{\rho^2+1}-\rho)\Big]\frac{d\hat{p}}{d\hat{h}}=0\, .
\ee
The expressions of $\hat{p}=\hat{p}(\hat{h})$ for the Hayward and Born-Infeld models are given in \eqref{QHH} and \eqref{QBI}, respectively.
For the derivatives of $\hat{p}$ over $\hat{h}$ one has $(1+\hat{h})^{-2}$ and $(1+\hat{h}^2)^{-3/2}$, respectively.
Substituting these expressions in \eqref{second}, one obtains the following relation
\be \n{hhrr1}
\begin{split}
&\hat{h}^{n}-c\hat{h}+k=0\, ,\\
&c=\dfrac{D-3}{2}\, ,\\
&k=\dfrac{D-1}{2}\beta \Big(
\sqrt{1+1/\rho^2}-1
\Big)\, .
\end{split}
\ee
It is remarkable that, in both cases, this equation has the same form. The only difference is the  parameter $n$, which takes the following values
\begin{itemize}
    \item $n = 2$ for the Hayward-type model;
    \item $n=3$ for the Born-Infeld-type model.
\end{itemize}
Let us note that for any number $D$ of spacetime dimensions one has that $c>0$. For $\rho\ge 0$ the function $k$ is positive and monotonically decreasing from $\infty$ at $\rho=0$ until it reaches $0$ at $\rho=\infty$.

We proceed as follows:
Suppose one solves the equation \eqref{hhrr1}, and let us denote by $H(\rho)$ its solution
\be
\hat{h}=H(\rho)\, .
\ee
Using the definition \eqref{hhhh} of $\hat{h}$ one concludes that the second extremality condition in \eqref{critical_conds} implies that
\be \n{HHH}
A\rho^{(1-D)/(D-2)}+\beta(F_D(\rho)-1)=H(\rho)\, .
\ee
Let $\rho=\rho_*$ denote the dimensionless radius at which the horizons coincide for an extremal black hole. One can then use \eqref{HHH} to determine the corresponding value of the parameter $A$
\begin{equation}\label{eq_Acr}
   A= \rho_*^{(D-1)/(D-2)}\Big[ H(\rho_*) - \beta \big( F_D(\rho_*) - 1 \big)\Big] \, .
\end{equation}
The value of the other dimensionless parameter $\hat{q}$, which enters the metric function $f$ given by \eqref{fff}, can then be determined. For this purpose, we use the first of Eqs.~\eqref{critical_conds}, which yields
\begin{equation}\n{qqqqq}
    \hat{q}^2=  \dfrac{1}{\rho_*^{2/(D-2)}\hat{p}(\rho_*)} \, .
\end{equation}
Here, $\hat{p}(\rho)$ denotes the function of $\rho$ obtained by substituting $\hat{h}=\hat{h}(\rho)$ into the model-dependent relation $\hat{p}=\hat{p}(\hat{h})$.

Thus, given the horizon radius $\rho_*$ of an extremally charged black hole, one can determine both parameters $A$ and $\hat{q}$ that enter the metric. These parameters, in turn, define the charge-to-mass ratio $\sigma$ and the dimensionless mass parameter $\hat{\mu}$ of the extremal solution
\be \n{sigsig}
\begin{split}
&\sigma=\dfrac{2\hat{Q}}{\mu} =\sqrt{\dfrac{2(D-1)}{(D-3)}}\sqrt{\beta}\frac{1}{\hat{q} A}\, , \\
&\hat{\mu} = A\hat{q}^{D-1}\ \, .
 \end{split}
 \end{equation}
By varying $\rho_*$, one obtains a parametric representation of $\sigma$ as a function of $\hat{\mu}$. In the next two subsections, we provide further details of this construction and present plots illustrating the extremality condition for charged black holes in the Hayward and Born–Infeld QTG models.

\subsection{Extremely charged black holes in the Hayward-type QTG}

\begin{figure}[!hbt]
    \centering
    \includegraphics[width=0.8
\linewidth]{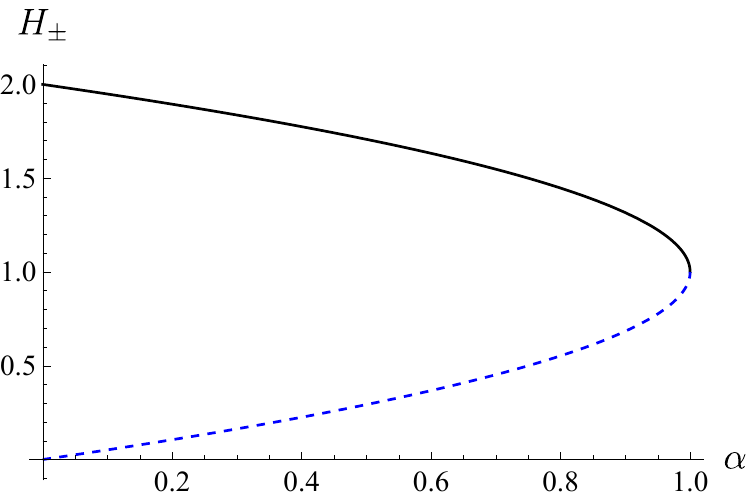}
    \caption{Plot of the functions $H_+$ (solid) and $H_-$ (dashed) defined in \eqref{SOL_HH}. Both solutions meet at $\alpha=1$. As $\alpha \to 0$, $H_+$ approaches the value $2$, while $H_-$ vanishes. }
    \label{FIG_HH}
\end{figure}

For the case of the Hayward-type QTG model, we consider the quadratic version of the equation \eqref{hhrr1}
\be
\hat{h}^{2}-c\hat{h}+k=0
\ee
Its solutions are
\be \label{SOL_HH}
\hat{h}_{\pm}=\dfrac{c}{2}H_{\pm}\hh
H_{\pm}=1
\pm \sqrt{1-\alpha}\hh \alpha=\dfrac{4k}{c^2} \, .
\ee
Thus, there always exists a point $\rho=\rho_m$ at which the expression inside the square root vanishes. For $\rho>\rho_m$, two solutions exist. One of them, $\hat{h}_-$, vanishes as $\rho \to \infty$, while the other, $\hat{h}_+$, approaches the value $c$. These two branches meet at $\rho=\rho_m$, where they share the common value $c/2$. This behavior is illustrated by the functions $H_\pm$ in Fig.~\ref{FIG_HH}.

\begin{figure}[!hbt]
    \centering
    \includegraphics[width=0.8
\linewidth]{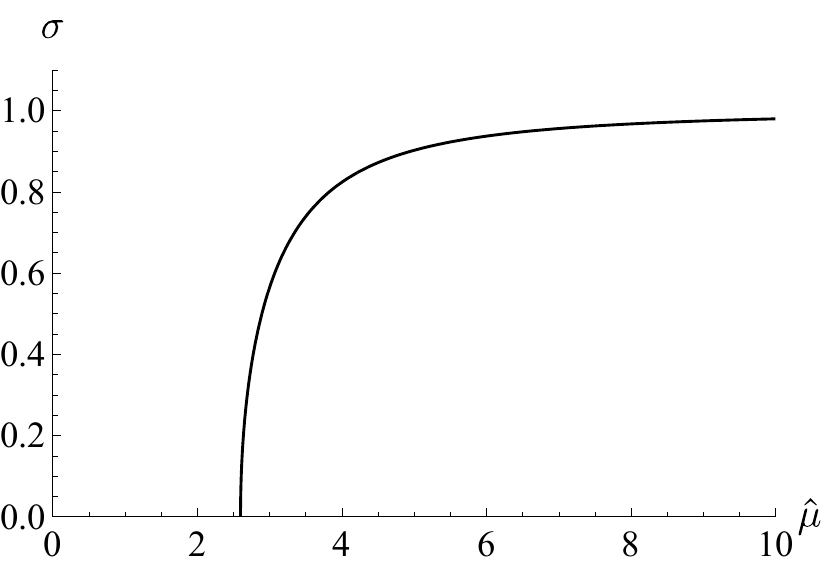}
    \caption{
Ratio $\sigma=2\hat{Q}/\mu$ for a critical charged black hole in the Hayward-like model, shown as a function of the dimensionless mass parameter $\hat{\mu}$. The curve $\sigma=\sigma(\hat{\mu})$ originates at $\hat{\mu}_* = 3\sqrt{3}/2$, where $\sigma=0$, corresponding to a neutral regular black hole of minimal mass. The plot is constructed for $\beta=1$ and $D=4$.}
    \label{FIG_sigH}
\end{figure}

The parametric plot for $\sigma(\hat{\mu})$, which describes the extremality of the black hole, is displayed on Figure \ref{FIG_sigH}

\subsection{Extremely charged black holes in the Born-Infeld-type QTG}

In this case, one has to consider the cubic version of \eqref{hhrr1}
\be \n{cubic}
\hat{h}^{3}-c\hat{h}+k=0\, .
\ee
Let us denote
\be
\Delta=\dfrac{k^2}{4}-\dfrac{c^3}{27}\, .
\ee
In the domain where $\Delta \ge 0$ the equation \eqref{cubic} has 3 real solutions. It is convenient to write these solutions in the following trigonometric form
\be \n{SOL_CUB}
\hat{h}_{i}=2\sqrt{\dfrac{c}{3}}H_i\hhh i=0,\pm\, .
\ee
Where
\be \n{SOL_H}
\begin{split}
&H_-=\cos\dfrac{\alpha +4\pi}{3}\, ,\\
&H_+=\cos\dfrac{\alpha }{3}\, ,\\
&H_0=\cos\dfrac{\alpha +2\pi}{3}\, ,\\
&\alpha=\arccos\Big(
-\dfrac{3\sqrt{3}k}{2c^{3/2}}
\Big)\, .
\end{split}
\ee
For a real solution, $k$ changes from $0$ (at $\rho=\infty$) to
\be
k_*=\dfrac{2c^{3/2}}{3\sqrt{3}}\, .
\ee

\begin{figure}[!hbt]
    \centering
    \includegraphics[width=0.8
\linewidth]{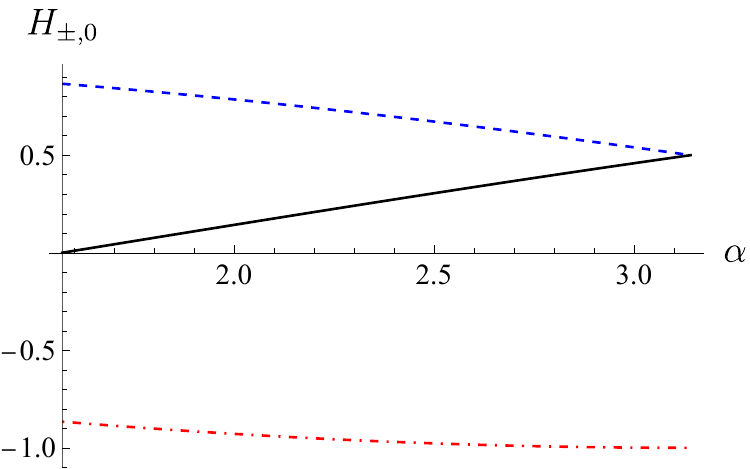}
    \caption{Plot of functions $H_-$ (solid), $H_+$ (dashed) and $H_0$ (dot-dashed) for $\alpha=[\pi/2,\pi]$.  }
    \label{FIG_HBI}
\end{figure}

To prove that \eqref{SOL_CUB}-\eqref{SOL_H} are really solutions of the cubic equation \eqref{cubic} it is sufficient to check that the following Vieta's relations are satisfied
\be
\begin{split}
&\hat{h}_-+\hat{h}_++\hat{h}_0=0\, ,\\
&\hat{h}_-\hat{h}_+
+\hat{h}_-\hat{h}_0
+\hat{h}_+\hat{h}_0=-c
\, ,\\
&\hat{h}_-\hat{h}_+\hat{h}_0=-k\, .
\end{split}
\ee
At $k=0$, corresponding to $\alpha=\pi/2$, one has $H_-=0$ and $H_+=\sqrt{3}/2$. For $k=k_*$, where $\alpha=\pi$ and $\Delta=0$, the two solutions $H_-$ and $H_+$ coincide and take the common value $1/2$. This behaviour is illustrated on Fig.~\ref{FIG_HBI}.

\begin{figure}[!hbt]
    \centering
    \includegraphics[width=0.8
\linewidth]{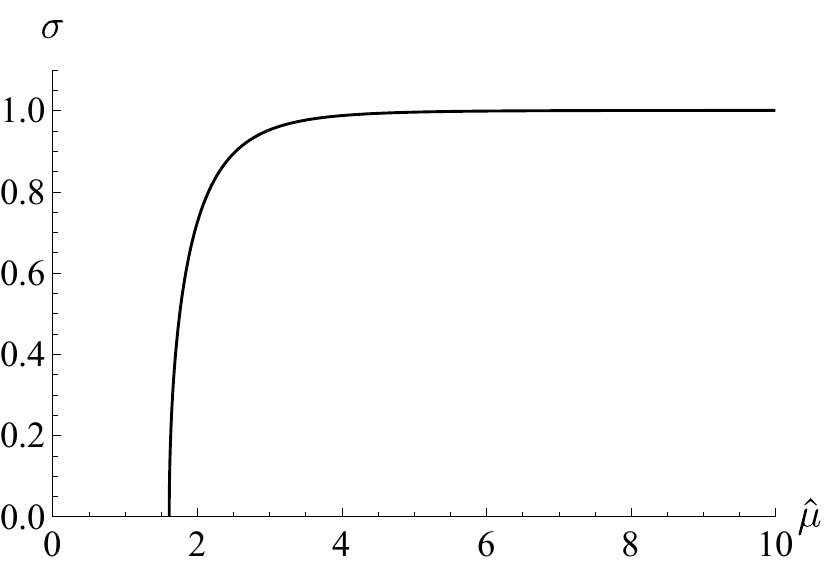}
    \caption{Ratio $\sigma=2\hat{Q}/\mu$ for a critical charged black hole in the Born–Infeld–type QTG model, shown as a function of the dimensionless mass parameter $\hat{\mu}$. The curve $\sigma=\sigma(\hat{\mu})$ originates at $\hat{\mu}_* = (3\sqrt{3}/2)^{1/2}$, where $\sigma=0$, corresponding to a neutral regular black hole of minimal mass. The plot is constructed for $\beta=1$ and $D=4$. }
    \label{FIG_sigBI}
\end{figure}

The parametric plot for the extremality parameter $\sigma(\hat{\mu})$ of this model is displayed on Figure~\ref{FIG_sigBI}

\section{Special cases}\label{sec_specialcases}

The model considered in this paper is characterized by two fundamental parameters: $\ell$, which sets the length scale at which QTG deviates from general relativity, and $b$, which determines the electric field strength at which nonlinear effects become significant. We now examine the solutions in two limiting cases, $\ell \to 0$ and $b \to \infty$. In the former limit, the theory reduces to Einstein gravity coupled to Born–Infeld nonlinear electrodynamics, while in the latter limit one recovers a Maxwell field coupled to QTG.

\subsection{Einstein-Born-Infeld solution}

We put $h(p)=p$ and keep the parameter $b$ finite. Then one has
\be \label{EBI_metric}
\begin{split}
&\CAL{E}= \frac{b}{\sqrt{1+\rho^2}}\, ,\\
&f=1-\dfrac{D-1}{D-2}\beta \hat{q}^2 \rho^{(3-D)/(D-2)}
\Big[
\hat{A}-J_D(\rho)\Big]\, .
\end{split}
\ee
Using relations \eqref{hhhh} and \eqref{ffqq}, one can verify that the parameters $\hat{A}$ and $\beta \hat{q}^2$, which enter the metric function $f$, are in fact independent of $\ell$, as expected. The metric function diverges at $\rho=0$.
Denote
\be
f_0=\dfrac{D-1}{D-2}\beta \hat{q}^2
\Big[\hat{A}-J_D^{(0)}\Big]\, ,
\ee
then at $\rho=0$ it has the following asymptotic form
\be \n{f00}
f=1-f_0 \rho^{(3-D)/(D-2)}+\ldots \, .
\ee
The metric function $f$ is singular at $\rho=0$, as expected, since the “smearing” effect associated with higher-curvature terms in the QTG model is absent. The metric possesses a $p$-sphere when $f_0<0$.

Restoring dimensions, the metric function $f$ can be written as follows:
\begin{equation}
\begin{split}
    f = 1 - \frac{\mu}{r^{D-3}} +&\frac{2 \varkappa b^2 r^2}{(D-2)(D-1)} \\
     & \bigg[1 - F_D\bigg(-\frac{Q^2}{b^2 r^{2D-4}} \bigg) \bigg]
\end{split}
\end{equation}

It is straightforward to verify that, at large distances, the metric asymptotically approaches the Reissner–Nordström-Tangherlini solution
\be
f|_{b\to \infty} = 1 - \frac{\mu}{r^{D-3}} + \frac{\hat{Q}^2}{r^{2(D-3)}} +\ldots\, .
\ee
This expression corresponds to the weakly nonlinear regime, in which Born–Infeld corrections appear only as small deviations from the Reissner–Nordström-Tangherlini solution. These nonlinear effects become significant only in the high-curvature region near the origin.

In $D=4$ dimensions one has
\begin{equation}
   f = 1 - \frac{\mu}{r} +\frac{  \varkappa b^2 r^2}{3}  \bigg[1 - F_4\bigg(-\frac{Q^2}{b^2 r^4} \bigg) \bigg]
\end{equation}
Using the identity \eqref{Identity_Hypergeom} derived in Appendix \ref{appxA}, one can show that the above equation reproduces the standard form of the metric for a spherically symmetric black hole in the Einstein–Born–Infeld theory\cite{hale2025rotextBI,GarciaD1984,Fernando2003,HdeOliveira_1994,Cai-BIADS,KUMARDEY2004484}.

\subsection{QTG-Maxwell models}

This model is obtained in the limit $b \to \infty$. Rather than taking this limit in the dimensionless form of the solution, it is more convenient to work directly with dimensionful quantities. In Sec.~\ref{Sec_RNT}, we have already derived the corresponding expression for $p$ for the Maxwell field. Using \eqref{pprr}, one obtains
\be
{h}=\dfrac{1}{r^{2(D-2)}}(
\mu r^{D-3} -\hat{Q}^2) \, .
\ee
Using these results, one can restore the primary curvature invariant $p$ and obtain the metric function via
\be
f(r)=1-r^2 p(r) \, .
\ee
The results are:
\begin{itemize}
\item For the Hayward-type QTG
\be \n{ppHH}
p(r)=\dfrac{\mu r^{D-3}-\hat{Q}^2}{r^{2(D-2)}+\ell^2 [\mu r^{D-3}-\hat{Q}^2]} \, .
\ee
\item For the Born-Infeld-type QTG
\be \n{ppBI}
p(r) =\dfrac{\mu r^{D-3}-\hat{Q}^2}{\sqrt{
r^{4(D-2)}+\ell^4 (\mu r^{D-3}-\hat{Q}^2)^2}
} \, .
\ee
\end{itemize}

For the Hayward case, the metric is singular at $r=r_0$, where $r_0$ is the root of the equation
\be
r^{2(D-2)}+\ell^2 [\mu r^{D-3}-\hat{Q}^2]=0\, .
\ee
For the Born-Infeld case, the metric is regular. These results are in agreement with similar approaches regarding charged black holes in QTG \cite{hao2026chargedregularblackholes}.

\subsection{Remarks on uncharged vacuum regular black holes in QTG}

The results obtained in the previous subsection can be readily specialized to the case of uncharged vacuum regular black holes in quasi-topological gravity by setting the electric charge to $\hat{Q}=0$. The corresponding solutions are well known and have been extensively discussed in the literature (see, e.g., \cite{QT_BH,Myers:2010ru,Bueno:2026dln,rbh_pfkz,Frolov:2026rcm,Moreno:2023rfl}).
For completeness, we briefly summarize these results here. As before, we express the metric functions in terms of the primary curvature invariant $p$
\be
f=1-r^2 p(r)\, .
\ee

\subsubsection{Uncharged vacuum black holes in the Hayward-type QTG }

Using expression \eqref{ppBI} and considering $\hat{Q}=0$, one gets
\be
p=\dfrac{\mu}{r^{D-1}+\ell^2\mu}\, .
\ee
At $r\to\infty$ one has that $p\sim \dfrac{\mu}{r^{D-1}}$, while at $r=0$ the curvature invariant goes to the constant value $p=p_0=1/\ell^2$. The equation $f=0$ gives
\be \n{HHhor}
r^{D-1} = \mu(r^2-\ell^2) \, .
\ee
Condition \eqref{HHhor} determines the locations of the horizons. There exists a critical value of the parameter, $\mu=\mu_*$. For $\mu<\mu_*$, Eq.~\eqref{HHhor} admits no real solutions, whereas for $\mu>\mu_*$ it has two solutions, corresponding to the inner and outer (event) horizons. The critical mass parameter $\mu_*$ is obtained by solving the system of two equations
\be \n{crit}
f=0\hh \dfrac{df}{dr}=0\, .
\ee
The second equation implies
\be \n{fp}
r^{D-1}=\dfrac{2\ell^2}{D-3} \mu\, .
\ee

Substituting this result in \eqref{HHhor} and solving the obtained equation one finds
\be \n{must}
\mu_*=\dfrac{1}{2}\dfrac{(D-1)^{(D-1)/2}}{(D-3)^{(D-3)/2}}\ell^{D-3}\, .
\ee
As shown earlier, the curve for the charge-to-mass ratio $\sigma$ vanishes at a certain value of the dimensionless mass parameter $\hat{\mu}=\mu/\ell^{D-3}$. At this point, the critical black hole is uncharged. Using \eqref{must}, one obtains the corresponding value $\hat{\mu}_*=\mu_*/\ell^{D-3}$, which is indicated in Fig.~\ref{FIG_sigH} at $\sigma=0$.

For $\mu>\mu_*$ the corresponding black hole is regular.
In the vicinity of the center $r=0$ the metric has a universal form
\be
f=1-\dfrac{r^2}{\ell^2}+\ldots \, .
\ee
This property is referred to as the existence of the deSitter core in the interior of a regular black hole.

\subsubsection{Uncharged vacuum black holes in the Born-Infeld-type QTG }

Using expression \eqref{ppBI} and setting $\hat{Q}=0$, one gets
\be
p=\dfrac{\mu}{\sqrt{r^{2(D-2)}+\ell^4\mu^2}}\, .
\ee
The equation defining the horizons, $f=0$ gives
\be \label{fpBI}
-\mu r^2+\sqrt{r^{2(D-1)}+\ell^4\mu^2}=0\, .
\ee
The second condition, $df/dr=0$, gives
\begin{equation}
    r^{2D-2} = \frac{2\ell^4\mu^2}{D-3} \, .
\end{equation}
Using this condition in \eqref{fpBI}, one can solve for the critical dimensionless mass parameter $\hat{\mu}_*$
\be \n{mustBI}
\hat{\mu}_*\equiv \dfrac{\mu_*}{\ell^{D-3}}=\dfrac{1}{\sqrt{2}}\dfrac{(D-1)^{(D-1)/4}}{(D-3)^{(D-3)/4}}\, .
\ee

This yields the expression for the critical mass parameter $\hat{\mu}_*$ corresponding to the point in Fig.~\ref{FIG_sigBI} where $\sigma=0$.
For $\hat{\mu}>\hat{\mu}_*$, the solution describes a regular black hole with two horizons and a de Sitter core, qualitatively similar to the Hayward-type black hole solution.

\section{Summary and Discussion}\label{sec7}

In this work, we have investigated static, spherically symmetric black hole solutions in quasi-topological gravity (QTG) coupled to nonlinear electrodynamics, with particular emphasis on the Born–Infeld model. Starting from the spherically reduced action, we showed that the field equations admit a universal representation in terms of the primary curvature invariant $p$ and two model-defining functions, $h(p)$ and $\mathcal{L}(\mathcal{E})$. This formulation enables one to construct solutions in integral form once the inverse relations $p=p(h)$ and $\mathcal{E}=\mathcal{E}(\mathcal{D})$ are known.
A key result of this work is the identification of a universal structure of charged solutions in QTG coupled to Born–Infeld electrodynamics. In dimensionless variables, the metric is determined by three parameters, $\beta$, $A$, and $\hat{q}$. The parameter $\beta$ characterizes the relative strength of higher-curvature and nonlinear electromagnetic effects. The remaining parameters, $A$ and $\hat{q}$, are defined in terms of the mass parameter $\mu$ and the charge $\hat{Q}$. This universality makes it possible to analyze the global properties of the solutions independently of the specific choice of the QTG model.

Our analysis shows that the interior structure of charged black holes in QTG crucially depends on the existence of a $p$-sphere, defined by the condition $h(p)=0$. This surface plays a central role in determining the regularity properties of the spacetime. In particular, when a $p$-sphere is present, the behavior of the solution inside it depends sensitively on the invertibility properties of the function $h(p)$.

We considered two representative classes of QTG models: the Hayward-type and the Born-Infeld-type. Although both models admit regular neutral black hole solutions, their charged counterparts exhibit qualitatively different behavior. In the Hayward-type model, generic charged solutions develop a curvature singularity at a finite radius inside the black hole. This can be interpreted as a shift of the singularity from the center to a finite-radius sphere, determined by the condition where the inverse function $p=p(h)$ ceases to be regular. Regular charged solutions still exist in this model, but only within a restricted range of parameters corresponding to sufficiently small masses.

In contrast, in the Born-Infeld-type QTG model, the function $p=p(h)$ remains regular for all values of $h$, and the corresponding charged black hole solutions are regular for all parameter values. The presence of a $p$-sphere does not lead to any pathology in this case; instead, the curvature invariants remain finite throughout the spacetime. Interestingly, the internal structure of these solutions differs qualitatively from that of neutral regular black holes: the de Sitter core characteristic of vacuum solutions is replaced by an anti-de Sitter core. This demonstrates that the inclusion of nonlinear electrodynamics can significantly modify the interior geometry, even when regularity is preserved.

We also analyzed extremal configurations and derived a general condition for extremality in terms of the dimensionless parameters of the solution. This condition reduces to an algebraic equation whose structure depends only on the type of QTG model under consideration. The resulting parametric representation of the extremality curve $\sigma=\sigma(\hat{\mu})$ provides a useful characterization of the allowed parameter space of charged black holes.

Finally, we examined several limiting regimes of the theory. In the limit $\ell \to 0$, the solutions reduce to those of Einstein gravity coupled to Born-Infeld electrodynamics, thereby recovering known results. In this case, black hole solutions are generically singular, except for special configurations in which the mass and charge parameters are fine-tuned.

In the opposite limit $b \to \infty$, the nonlinear electrodynamics reduces to Maxwell theory, and one obtains charged black hole solutions in pure QTG. In this case, the regularizing effect of nonlinear electrodynamics is absent. However, the Born-Infeld-type QTG model still admits regular black hole solutions.

In summary, our results demonstrate that quasi-topological gravity provides a flexible framework for constructing regular black hole solutions, but the inclusion of charge introduces qualitatively new features. The interplay between higher-curvature corrections and nonlinear electrodynamics determines whether the resulting spacetime remains regular or develops singularities. The Born-Infeld-type QTG model appears to be particularly robust in this respect, yielding regular charged black holes for arbitrary parameters. These findings may be relevant for understanding the internal structure of black holes in effective theories of gravity and for exploring scenarios in which singularities are resolved.

\appendix

\section{Hypergeometric Function Identity}\label{appxA}

In this appendix, we derive a relation between the hypergeometric function with different arguments, which is used in section~\ref{sec_specialcases}.
Consider the Euler integral representation of the hypergeometric function \cite{NIST2010}
\begin{equation}\label{eulerInt}
\begin{split}
    &{}_2F_1(a,b;c;z) = \\
    & \frac{\Gamma(c)}{\Gamma(b)\Gamma(c-b)} \int_0^1 t^{b-1}(1-t)^{c-b-1}(1-zt)^{-a}dt \, .
\end{split}
\end{equation}
We are interested in a particular case, specifically when $a=-1/2$ and $c = b+1$.
For these parameters the relation \eqref{eulerInt} takes the form
\begin{equation}
     {}_2F_1(-1/2,b;b+1;z) = b\int_0^1 t^{b-1}(1-zt)^{1/2}dt \, .
\end{equation}
By performing an integration by parts, one obtains
\begin{equation}\n{int}
\begin{split}
    b\int_0^1 t^{b-1}&(1-zt)^{1/2}dt = (1-z)^{1/2} +\\
    & \frac{z}{2}\int_0^1 t^b(1-zt)^{-1/2}dt \, .
\end{split}
\end{equation}
The integral in the right-hand side can be expressed again in terms of the hypergeometric function
\begin{equation}
\int_0^1 t^b(1-zt)^{-1/2}dt=\dfrac{1}{b+1} {}_2F_1(1/2,b+1;b+2;z)   \, .
\end{equation}
Therefore
\begin{equation}\label{Identity_Hypergeom}
\begin{split}
   {}_2F_1&(-1/2,b;b+1;z)  = (1-z)^{1/2} \\
   &+ \frac{z}{2(b+1)} {}_2F_1(1/2,b+1;b+2;z) \, .
\end{split}
\end{equation}
This is the identity used in the main body of the paper.

\section*{Acknowledgments}

The authors thank the Natural Sciences and Engineering Research Council of Canada for support. V.F. is also supported by the Killam Trust. During his stay at Nagoya University, V.F. acknowledges support from the Institute for Advanced Research (IAR) through the International PI Invitation Program and thanks Prof. Akihiro Ishibashi and Dr. Chulmoon Yoo for their hospitality and stimulating discussions.

\bibliography{references}

\end{document}